\documentclass[aps,epsf,subfigure,twocolumn]{revtex4}

\usepackage{graphicx}% Include figure files
\usepackage{epstopdf}
\usepackage{dcolumn}% Align table columns on decimal point
\usepackage{bm}% bold math
\usepackage{amsmath}
\usepackage{color,psfrag,epsfig}
\begin{document}

\newcommand{\bk}{{\bf k}}
\newcommand{\bc}{\begin{center}}
\newcommand{\ec}{\end{center}}
\newcommand{\mub}{{\mu_{\rm B}}}
\newcommand{\sD}{{\scriptscriptstyle D}}
\newcommand{\sF}{{\scriptscriptstyle F}}
\newcommand{\sCF}{{\scriptscriptstyle \mathrm{CF}}}
\newcommand{\sH}{{\scriptscriptstyle H}}
\newcommand{\sAL}{{\scriptscriptstyle \mathrm{AL}}}
\newcommand{\sMT}{{\scriptscriptstyle \mathrm{MT}}}
\newcommand{\sT}{{\scriptscriptstyle T}}
\newcommand{\up}{{\mid \uparrow \rangle}}
\newcommand{\down}{{\mid \downarrow \rangle}}
\newcommand{\upsp}{{\mid \uparrow_s \rangle}}
\newcommand{\downsp}{{\mid \downarrow_s \rangle}}
\newcommand{\upsone}{{\mid \uparrow_{s-1} \rangle}}
\newcommand{\downsone}{{\mid \downarrow_{s-1} \rangle}}
\newcommand{\upt}{{ \langle \uparrow \mid}}
\newcommand{\downt}{{\langle \downarrow \mid}}
\newcommand{\bbar}{{\mid \uparrow, 7/2 \rangle}}
\newcommand{\abar}{{\mid \downarrow, 7/2 \rangle}}
\renewcommand{\a}{{\mid \uparrow, -7/2 \rangle}}
\renewcommand{\b}{{\mid \downarrow, -7/2 \rangle}}
\newcommand{\plus}{{\mid + \rangle}}
\newcommand{\minus}{{\mid - \rangle}}
\newcommand{\psio}{{\mid \psi_o \rangle}}
\newcommand{\psis}{{\mid \psi \rangle}}
\newcommand{\bpsio}{{\langle \psi_o \mid}}
\newcommand{\barpsi}{{\mid \psi' \rangle}}
\newcommand{\barpsio}{{\mid \bar{\psi_o} \rangle}}
\newcommand{\ex}{{\mid \Gamma_2^l \rangle}}
\newcommand{\LH}{{{\rm LiHoF_4}}}
\newcommand{\LHx}{{{\rm LiHo_xY_{1-x}F_4}}}
\newcommand{\de}{{{\delta E}}}
\newcommand{\Ht}{{{H_t}}}

\title{Identification of strong and weak interacting two level systems in KBr:CN}

\author{Alejandro Gaita-Ari\~no$^1$}
\author{Moshe Schechter$^2$}
\affiliation{$^1$Department of Physics \& Astronomy, University of British Columbia,
Vancouver, B.C., Canada, V6T 1Z1}
\affiliation{$^2$Department of Physics, Ben Gurion University - Beer Sheva 84105, Israel}
\date{\today}

\begin{abstract}

Tunneling two level systems (TLSs) are believed to be the source of phenomena
such as the universal low temperature properties in disordered and amorphous
solids, and $1/f$ noise. The existence of these phenomena in a large
variety of dissimilar physical systems testifies for the universal nature of
the TLSs, which however, is not yet known. Following a recent suggestion that
attributes the low temperature TLSs to inversion pairs [M. Schechter and 
P. C. E. Stamp, arXiv:0910.1283.] we calculate explicitly the TLS-phonon 
coupling of inversion symmetric and asymmetric TLSs in a given disordered 
crystal. Our work (a) estimates parameters that support the theory in 
M. Schechter and P. C. E. Stamp, arXiv:0910.1283, in its general form, and (b) 
positively identifies, for the first time, the relevant TLSs in a given system.

\end{abstract}

\maketitle

{\it Introduction} ---
Amorphous solids and many disordered lattices show peculiar universal characteristics at low
temperatures\cite{ZP71,HR86,PLT02}. Below $T_U \approx 3$K systems which are
otherwise very different have specific heat $C_v \propto T^{\alpha}$, with
$\alpha \approx 1$, thermal conductivity $\kappa \propto T^{\beta}$ with $\beta
\approx 2$, and internal friction $Q \approx 2 \pi l/\lambda \approx 10^3$,
independent of $T$, $\lambda$, and with only a small variance between materials.
Here $l$ is the phonon mean free path and $\lambda$ is the phonon wavelength. In an
effort to explain this remarkable universality, Anderson Halperin and
Varma\cite{AHV72} and Philips\cite{Phi72} suggested a phenomenological theory,
where the existence of tunneling two level systems (TLSs) in these materials
was postulated, and an ansatz for their density of states was given.  This
"standard tunneling model" (STM) has been very successful in explaining
the above mentioned phenomena. Still, the identity of the tunneling TLSs has
remained unknown.
Furthermore, the smallness and universality of the phonon attenuation, and the
energy scale dictating $T_U$ are not accounted for by the STM.

Two level systems are also believed to be the cause of $1/f$ noise. Recently,
it has been shown that $1/f$ noise is the main source for decoherence of
superconducting qubits, and a major obstacle in their ability to perform quantum
computation\cite{SLH+04}. Also in these systems the nature of the TLSs is not known, yet
assuming their existence and applying the STM has resulted in an explanation of
the low frequency $1/f$ noise and high frequency linear in $f$ noise on the
same footing\cite{SSMM05}.

Extensive experimental investigations have revealed that the condition to
observe universality is the presence of tunneling states and strong lattice
strain\cite{Wat95,TTP99}, and that the phenomena in amorphous solids and disordered
crystals are equivalent\cite{LVP+98}. Disordered crystals are advantageous for
both experimental and theoretical investigation\cite{PLC99}. Experimentally,
they allow control of the nature and relative concentration of host
material and impurities, and therefore a detailed study of different universal
properties and their origin. The existence of lattice structure and the
apparent candidates for tunneling states allows a favorable starting point for
theoretical treatment as well.

Indeed, it was argued\cite{SS08b} that, at least in disordered crystals,
tunneling states can be categorized into two types of TLSs, denoted $\tau$ and
$S$. The states of a $\tau$-TLS are related to each other by inversion.
Consequently, the interaction of a $\tau$-TLS with the phonon field
$\gamma_{\rm w}$ is small, as it results only from disorder induced local
deviations from inversion symmetry. The $S$-TLSs are asymmetric with
respect to local inversion, with a strong interaction with the phonon field
$\gamma_{\rm s}$. It was then shown\cite{SS08b} that the $S$-TLSs are gapped below $T_U$ by
the $\tau$-TLSs through an Efros Shklovskii\cite{ES75} type mechanism, and
that below $T_U$ the $\tau$-TLSs are effectively noninteracting, and
dictate the phonon attenuation.
Thus, at $T < T_U$ the $\tau$-TLSs fulfil the assumptions of the STM. The small
parameter of the theory is $g \equiv \gamma_{\rm w}/\gamma_{\rm s} \approx
E_{\phi}/E_C \sim (1-3) \times 10^{-2}$, where $E_{\phi}, E_C$ are the typical
elastic and Coulomb energies in the system.  This small parameter gives the
universality and smallness of the phonon attenuation. Defining $T_G$ as the
ordering temperature of the $S$-TLSs, the emerging DOS of the $S$-TLSs at an
energy $T_U \approx g T_G$, dictates $T_U$ as the energy scale below which
universality is observed\cite{SS08b}.

\begin{figure}[htb]
\begin{tabular}{ccccc}
\hspace{0.90cm}
\raisebox{-0.3cm}[0.0cm][0.0cm]{(a)}
\hspace{-1.65cm}
\includegraphics[angle=00.0,width=0.14\textwidth]{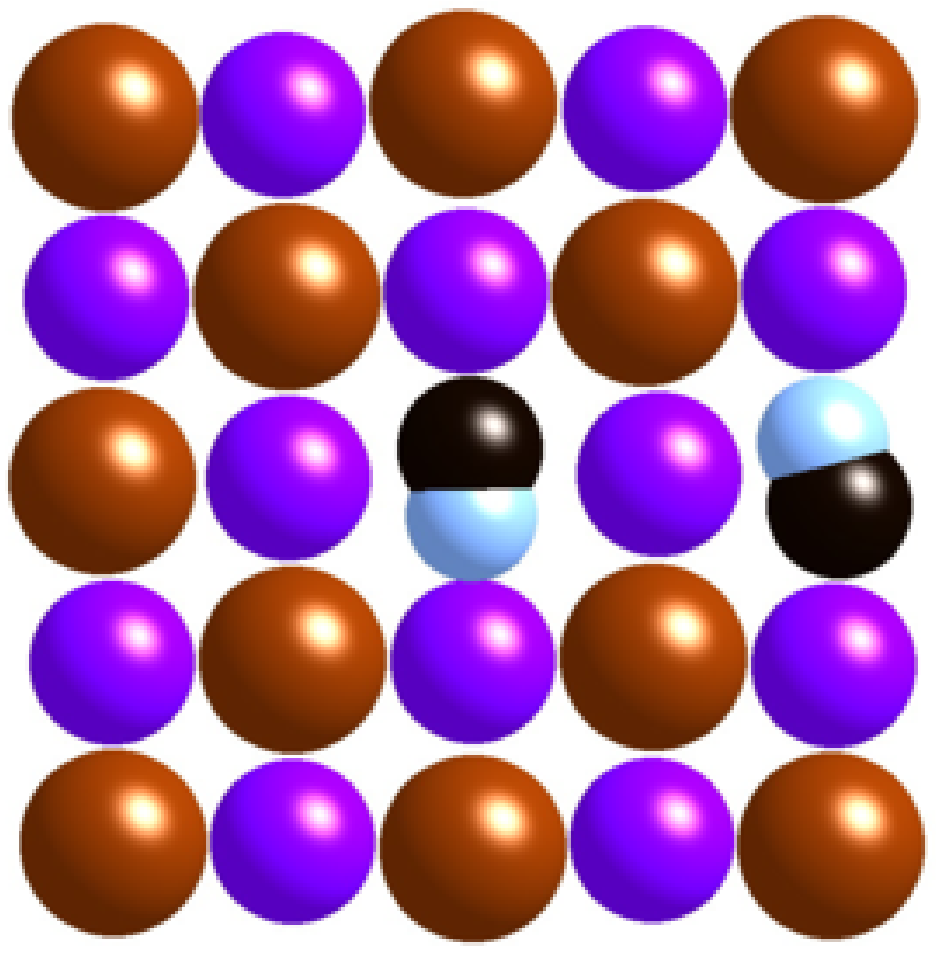} &
\hspace{-0.20cm}
\raisebox{1.40cm}{$\tau_{\rm ud}$} \hspace{-0.70cm}
\raisebox{1.10cm}{$\longleftarrow$} &
\hspace{0.75cm}
\raisebox{-0.3cm}[0.0cm][0.0cm]{(b)}
\hspace{-1.65cm}
\includegraphics[angle=00.0,width=0.14\textwidth]{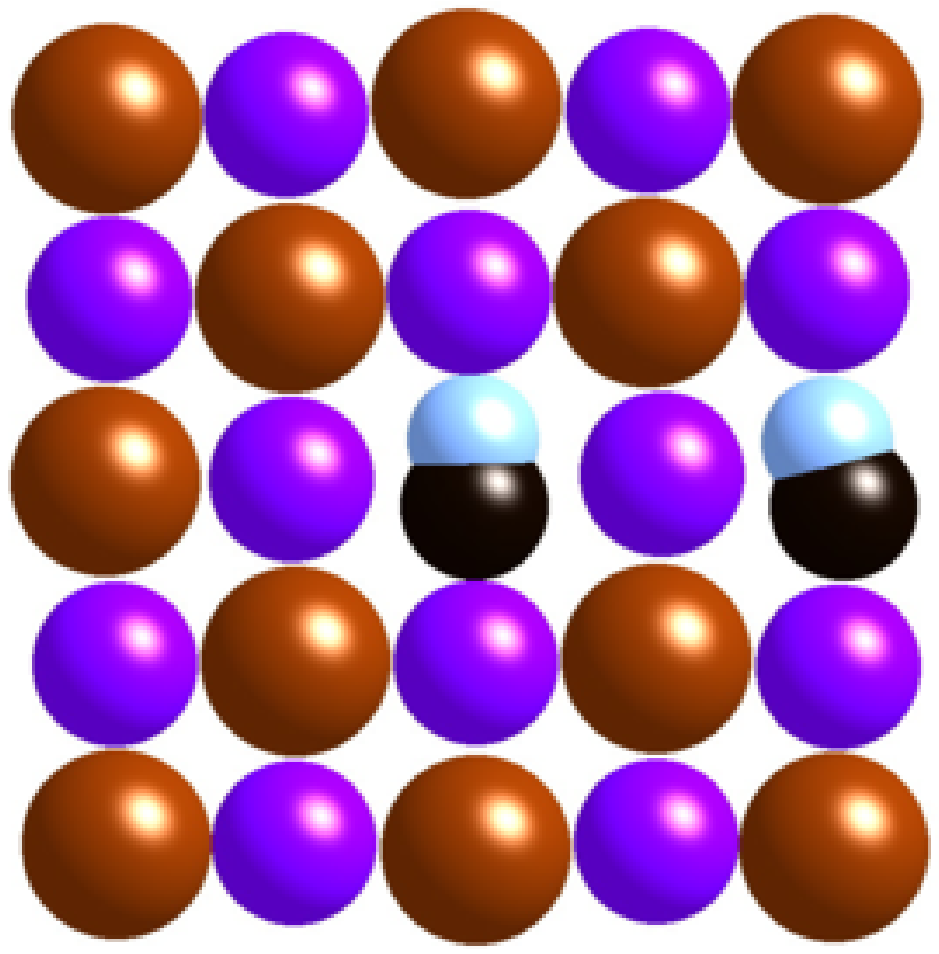} &
\hspace{-0.25cm}
\raisebox{1.40cm}{$S_{\rm ul}$} \hspace{-0.70cm}
\raisebox{1.10cm}{$\longrightarrow$} &
\hspace{0.90cm}
\raisebox{-0.3cm}[0.0cm][0.0cm]{(c)}
\hspace{-1.65cm}
\includegraphics[angle=00.0,width=0.14\textwidth]{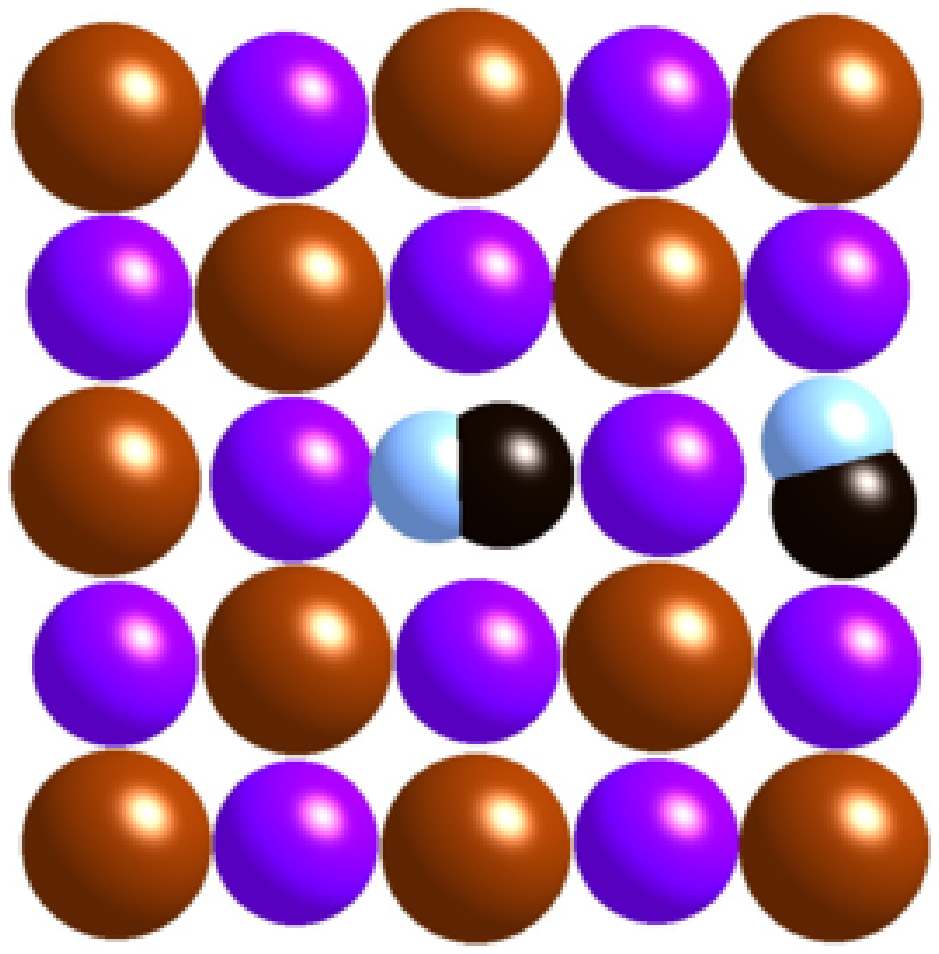} \\
\end{tabular}
\caption{Three 5x5 fragments of a KBr:CN lattice. The \emph{up} state of
the central impurity in fragment (b) is related by a $\tau$ excitation to a
\emph{down} state in fragment (a), and by an $S$ excitation to a \emph{left}
state in fragment (c).}
\label{TLS}
\end{figure}

In this Letter we use DFT and ab-initio calculations
to calculate the interaction of TLSs of types $\tau$ and $S$ with the phonon
field in the system KBr$_{(1-x)}$(CN)$_x$ (KBr:CN, Fig.~\ref{TLS}). We find
that $\gamma_{\rm w} \approx 0.1$eV, and $\gamma_{\rm s}
\approx 3$eV. Our estimation of $\gamma_{\rm w}$ compares well with the experimentally measured
value for the relevant TLSs at low energies, of $\gamma \approx 0.12$eV for
impurity concentration x$= 0.25$ and $\gamma \approx 0.2$eV for x$=
0.5$\cite{BDL+85,YKMP86}. Our results also support the central arguments of the
theory in Ref.\cite{SS08b} in (i) the categorization
of the TLSs according to their symmetry under inversion, (ii) the ratio of the
strengths of their interactions with the phonon field, constituting the small
parameter of the theory, and (iii) the identification of the symmetric TLSs as the
relevant TLSs dictating the low temperature universal properties in disordered
solids.
In addition, we re-enforce the prediction made in Ref.\cite{SS08b} for the
existence, at higher energies, of a second type of TLSs (of $S$ type), with a
much stronger coupling to the phonon field. Note, that although we focus here on the simplest single impurity excitations, our analysis does not exclude the possibility of symmetric and asymmetric multi-impurity excitations\cite{GRS90,Bur95,BK96}. Such excitations are expected to be significant especially for systems where single impurity excitations do not produce symmetric TLSs\cite{NYHC87,YNH89}.

For the specific KBr:CN system, early works have suggested, based on theories quite different from that of Ref.\cite{SS08b}, that CN flips comprise the relevant low energy TLSs\cite{SC85,SK94}. However, for long, advance in this direction was hindered because of experiments showing that the substitution of the symmetric N$_2$ molecules for the asymmetric CO molecules in N$_2$/Ar/CO does not change its universal characteristics\cite{NYHC87,YNH89}. Our results here, in conjunction with the theory in Ref.\cite{SS08b}, positively identify the $180^\circ$ CN flips as the relevant TLS excitations dictating the low temperature characteristics in the KBr:CN system.
Reconciliation of our results with the experiment in Ref.\cite{NYHC87,YNH89} stems from the fact that pairs of N$_2$ molecules do produce symmetric TLSs in the ArN$_2$ system\cite{GAS11}.

{\it Calculation} ---
KBr:CN is perhaps the most studied disordered lattice showing universal characteristics.
The CN$^-$ impurities have been found to orient either in the direction of the
in-space diagonals, preferred for very low CN$^-$
concentrations\cite{Bey75} and for intermediate concentrations at high
temperatures\cite{LKRM88}, or in the direction of the axes,
preferred for intermediate CN$^-$ concentrations at low temperatures\cite{LKRM88}.
The six (eight) possible states of each impurity can be categorized
into three (four) inversion pairs, each having two states related to each other
by an $180^{\circ}$ flip.  Such flips constitute $\tau$ excitations, whereas
rotations between different axis (diagonals) correspond to $S$
excitations\cite{SS08a,SS08b}.

The interaction of such a system with the lattice can be described by the Hamiltonian\cite{SS08a,SS08b}
\begin{equation}
H_{\rm int} \;=\; \sum_j \sum_{\alpha,\beta} \;
[\eta\delta^{\alpha,\beta} + \gamma_{\rm s}^{\alpha\beta} S_j^z \;+\;
\gamma_{\rm w}^{\alpha\beta} \tau_j^z] \; u_{\alpha\beta}({\bf r}_j)
\label{Hint}
\end{equation}
where $\eta$ is an orientation-independent volume factor and $u_{\alpha\beta}({\bf r}_j)$
denotes the phonon field at point ${\bf r}_j$.
Whereas the central purpose of this Letter is the calculation of $\gamma_{\rm w}$
and $\gamma_{\rm s}$, we also calculate the parameter $\eta$ for both CN$^-$ and Cl$^-$
impurities. This parameter determines the strain, and thus the effective random
field in the system\cite{SS08a,SS09}. Usually $\eta \lesssim \gamma_{\rm s}$.
In KBr:CN $\eta$ is significantly subdominant, as the Br$^-$ and CN$^-$ ions
have similar volumes\cite{YKMP86}. In KBr:Cl this term is responsible for the
strains allowing for the existence of universal properties upon minimal CN$^-$
dilution\cite{Wat95,TTP99}. This random field
term is also central to the smearing of the glass transition and the peculiar
disordering of dilute glasses\cite{SS09}.

Following the above definition of $\eta$, $\gamma_{\rm s}$ and  $\gamma_{\rm
w}$, we devise a series of numerical calculations to estimate them. In sum, we
choose a number of lattice fragments and use DFT/ab initio methods to calculate
the energy difference between the effects of phonon-like perturbations of the
system with a central CN$^-$ impurity in different states.  For simplicity and
without loss of generality, we restrict the excitations of the CN$^-$ impurity
under study to two dimensions, thus the possible states are \emph{up},
\emph{down}, \emph{left} and \emph{right}. To those possible
orientations, we apply \emph{vertical} or \emph{horizontal} phonons. Usually
the symmetry is low enough to allow for several independent estimations for
each parameter.

Our determination of $\gamma_{\rm w}$ and $\gamma_{\rm s}$ is performed as follows:
\begin{equation}
\gamma_{\rm w,s}=\frac{1}{b}\cdot|(E^i_{ph}(b)-E^i)-(E^j_{ph}(b)-E^j)|
\end{equation}
where \{i, j\} are \{up, down\} or \{left, right\} for $\gamma_{\rm w}$ and
\{up, left\} or \{down, right\} for $\gamma_{\rm s}$,
$ph$ can stand for $vertical$ or $horizontal$ phonons, $E^i$ is the energy of
an impurity $i$ surrounded by a lattice fragment in its equilibrium geometry
and $E^i_{ph}$ is the energy of the same impurity after a lattice contraction
by a fraction $b$ along a given crystallographic coordinate, mimicking the
effect of a longitudinal phonon.
For $\eta$, the same procedure is
applied where \{i,j\} means presence or absence of impurity, and for CN$^-$
impurities all possible orientations are averaged.

\begin{figure}[htb]
\begin{tabular}{ccc}
\hspace{0.90cm}
\raisebox{-0.3cm}[0.0cm][0.0cm]{(a)}
\hspace{-1.80cm}
\includegraphics[angle=00.0,width=0.158\textwidth]{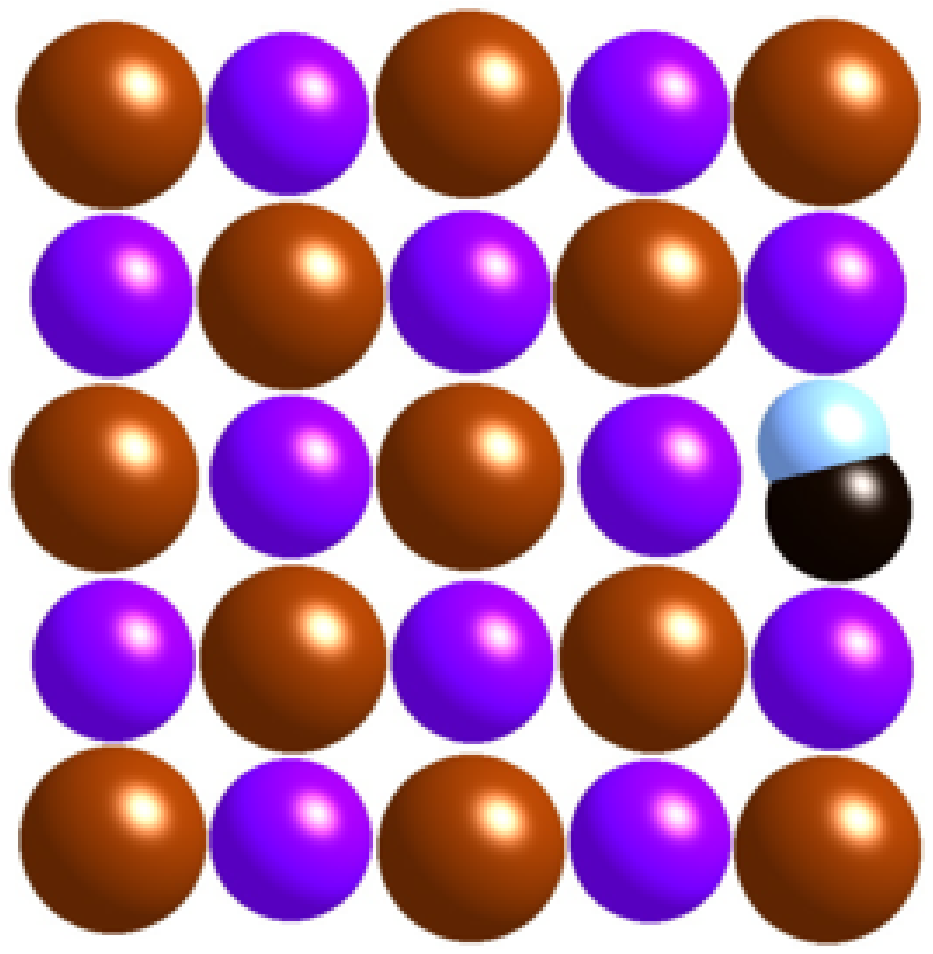} &
\hspace{-0.20cm}
\hspace{1.25cm}
\raisebox{-0.3cm}[0.0cm][0.0cm]{(b)}
\hspace{-1.80cm}
\includegraphics[angle=01.4,width=0.155\textwidth]{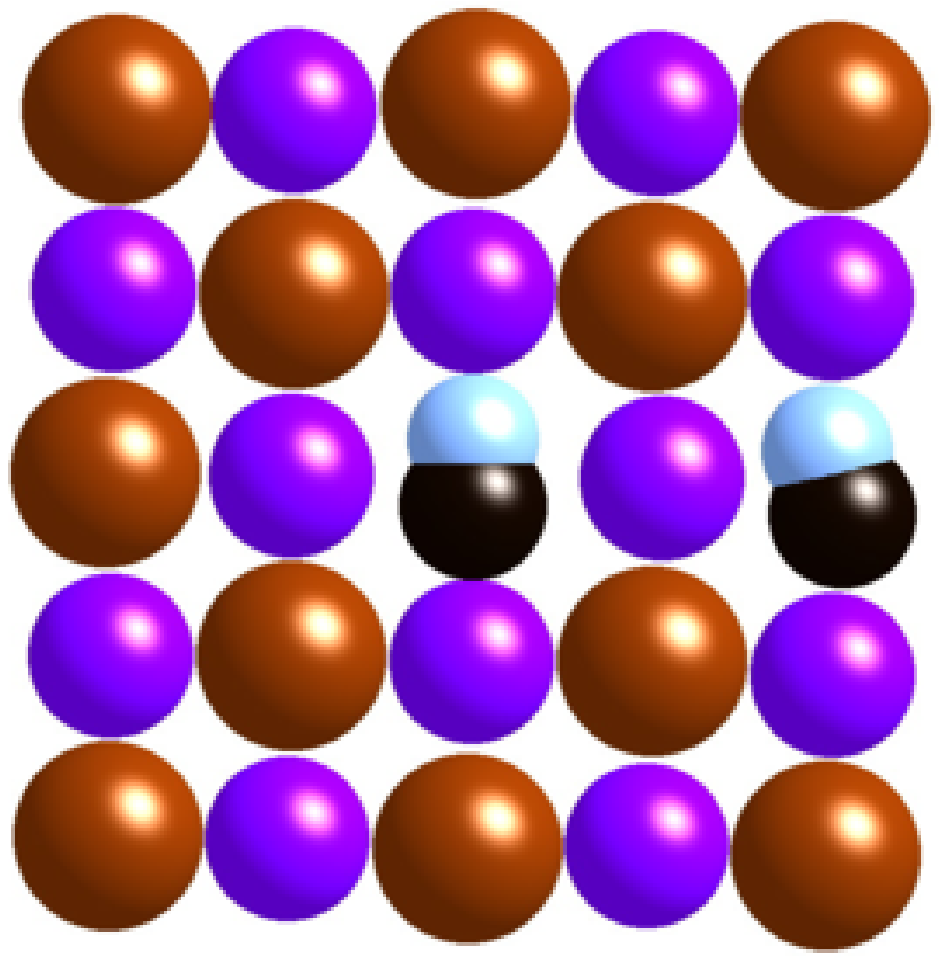} &
\hspace{-0.25cm}
\hspace{1.25cm}
\raisebox{-0.3cm}[0.0cm][0.0cm]{(c)}
\hspace{-1.80cm}
\includegraphics[angle=00.0,width=0.156\textwidth]{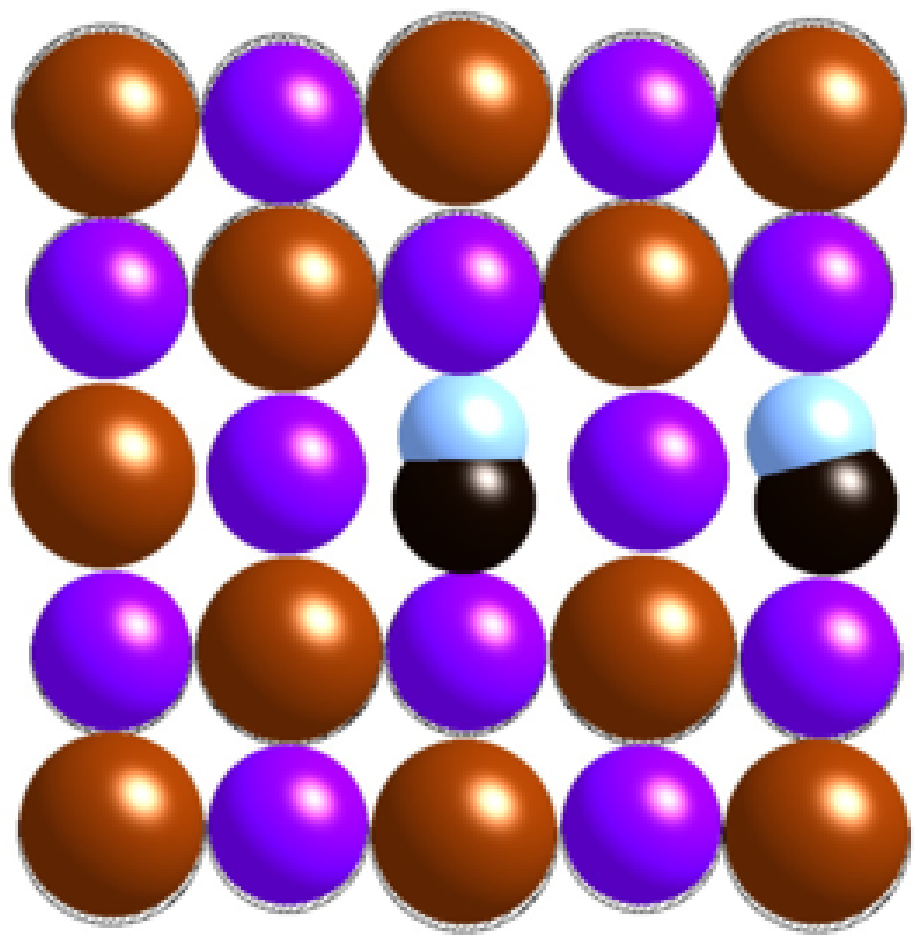} \\
\end{tabular}
\caption{Atomic coordinates of a lattice fragment in three key steps of a
sample calculation. (a): all positions are optimized, without a central TLS,
but in presence of an impurity (2,0,u). (b): a TLS (0,0,u) is substituted in
the place of the central Br$^-$, its position is optimized, yielding $E^i$.
(c): the lattice is contracted by 2\% vertically, yielding
$E^i_{vertical}(-2\%)$.} \label{relax}
\end{figure}

\begin{figure}[htp]
\vspace{0.0cm}\hspace{-0.2cm}
\includegraphics[width=0.42\textwidth]{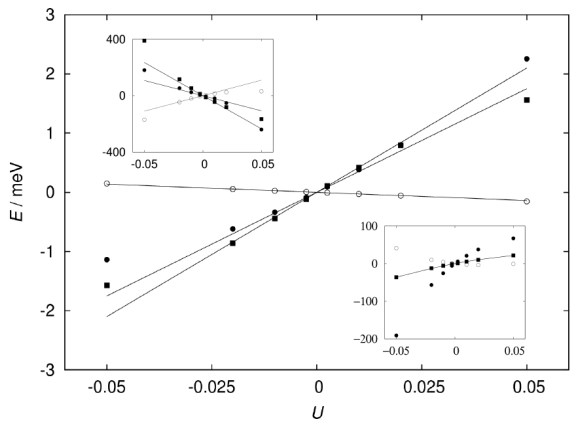}
\caption{Energy response $E$ (in meV) to horizontal phonons of amplitude $U$.
$\gamma_{\rm w}$ (main graph) and $\gamma_{\rm s}$ (upper inset) as linear fits
on selected fragments on table~\ref{tablegamma}: open circles: (b); filled
circles: (c); squares: (d).  Lower inset: second-order fit for $\eta$ (see text);
open circles: $E_{Br}-E_{up}$; filled circles: $E_{Br}-E_{left}$; squares:
weighted average for all orientations.}
\label{relaxnewfigure}
\end{figure}

Fig.~\ref{relax} illustrates the detailed procedure of a $\gamma$ calculation.
We describe the position and orientation of the impurity as (2,0,u), where the
numbers refer to the coordinates and the letter to the orientation, in this
case the CN$^-$ is lying on the abscise, at 2 interatomic spacings to the
right, and the nitrogen is pointing up. We then proceed with the following
steps: (i) for that particular system, the atomic positions are found
which minimize the energy in the absence of the TLS under investigation, i.e.,
when the central position is occupied by a Br$^-$; that is our definition of
the equilibrium geometry of the lattice. (ii) a CN$^-$ (facing "up", in
this case) is substituted for the central Br$^-$, and, freezing the lattice,
only the position of the CN$^-$ is optimized, to obtain $E^{up}$. (iii) the
atomic positions are contracted along the vertical axis to obtain
$E^{up}_{vertical}(b)$.
The calculation is done for $b\in\{-5\%,
-2\%, -1\%, -0.25\%, 0.25\%, 1\%, 2\%, 5\%\}$, and the limit of small $b$ is
taken (see Fig.~\ref{relaxnewfigure}).
The repetition of this procedure for the vertical
phonon and the "down", "left" and "right" orientations of the central CN$^-$
allows for four non-independent estimations of both $\gamma_{\rm w}$ and
$\gamma_{\rm s}$ (in 3D we also have the "front" and "back" orientations). In
the case of $\eta$, the weighted average for all orientations yields one unique
estimation. Note that in step (ii) we do not relax the whole lattice in the presence
of the TLS. Our procedure is in line with both processes of lattice relaxation and
phonon scattering by TLSs, which result from the out of equilibrium first order interaction
of the TLS with the lattice.

We use the standard package Gaussian03~\cite{gaussian03} to perform quantum
chemistry calculations on lattice fragments of different sizes and shapes and
at different levels of sophistication. As we are dealing with a local
phenomenon, and for cost reasons, most of the calculations are performed on
small zero-dimensional squares or cubes, either 3x3, 3x3x3 or 5x5.  The TLS
under evaluation is always in the center, so that any deviation from a
centrosymmetric situation felt by the TLS is due to the extra impurities and
not to border effects.  We mainly use the hybrid DFT/\emph{ab initio} method
B3LYP with small orbital sets, either 3-21G or 6-31G. The influence of a better
description of the anions is tested by repeating some calculations with the
more flexible basis sets, up to 6-311+G*.  Additionally, we check the relevance
of dinamical correlation by comparing plain Hartree-Fock with the
Moeller-Plesset perturbation theory to the second order (MP2), which includes
double excitations as second-order perturbations.
Last, we include a limited study of larger samples with a higher number of
impurities, using HFS calculations with the minimal basis STO-3G on a 7x7
fragment with up to 4 extra impurities.

\begin{table}[htb]
\vspace{0.3cm}
\begin{tabular}{|c|c|c|c|c|}
\hline
\raisebox{-0.1cm}[0.0cm][0.25cm]{fragment} & \raisebox{-0.1cm}[0.0cm][0.25cm]{impurity} & \raisebox{-0.1cm}[0.0cm][0.25cm]{method} & \raisebox{-0.1cm}[0.0cm][0.25cm]{$\gamma_s$}   &  \raisebox{-0.1cm}[0.0cm][0.25cm]{$\gamma_w$}   \\
\hline
\hline
\raisebox{-0.05cm}[0.0cm][0.15cm]{\scriptsize{3x3}}    & \raisebox{-0.05cm}[0.0cm][0.15cm]{\scriptsize{none}}     & \raisebox{-0.05cm}[0.0cm][0.15cm]{\scriptsize{HF/6-311+G*}} &    \raisebox{-0.05cm}[0.0cm][0.15cm]{\scriptsize{$3.02$}}    &     \raisebox{-0.05cm}[0.0cm][0.15cm]{\scriptsize{0 (sym)}}      \\
\hline
\raisebox{-0.05cm}[0.0cm][0.15cm]{\scriptsize{3x3}}    & \raisebox{-0.05cm}[0.0cm][0.15cm]{\scriptsize{none}}     & \raisebox{-0.05cm}[0.0cm][0.15cm]{\scriptsize{MP2/3-21G}} &    \raisebox{-0.05cm}[0.0cm][0.15cm]{\scriptsize{$1.91$}}    &     \raisebox{-0.05cm}[0.0cm][0.15cm]{\scriptsize{0 (sym)}}      \\
\hline
\raisebox{-0.05cm}[0.0cm][0.15cm]{\scriptsize{3x3}}    & \raisebox{-0.05cm}[0.0cm][0.15cm]{\scriptsize{none}}     & \raisebox{-0.05cm}[0.0cm][0.15cm]{\scriptsize{MP2/6-311+G*}} &    \raisebox{-0.05cm}[0.0cm][0.15cm]{\scriptsize{$3.55$}}    &     \raisebox{-0.05cm}[0.0cm][0.15cm]{\scriptsize{0 (sym)}}      \\
\hline
\raisebox{-0.05cm}[0.0cm][0.15cm]{\scriptsize{3x3}}    & \raisebox{-0.05cm}[0.0cm][0.15cm]{\scriptsize{none}}     & \raisebox{-0.05cm}[0.0cm][0.15cm]{\scriptsize{B3LYP/3-21G}} &    \raisebox{-0.05cm}[0.0cm][0.15cm]{\scriptsize{$3.28$}}    &     \raisebox{-0.05cm}[0.0cm][0.15cm]{\scriptsize{0 (sym)}}      \\
\hline
\raisebox{-0.05cm}[0.0cm][0.15cm]{\scriptsize{3x3 (tr.ph.)}}    & \raisebox{-0.05cm}[0.0cm][0.15cm]{\scriptsize{none}}     & \raisebox{-0.05cm}[0.0cm][0.15cm]{\scriptsize{B3LYP/3-21G}} &    \raisebox{-0.05cm}[0.0cm][0.15cm]{\scriptsize{$2.95$}}    &     \raisebox{-0.05cm}[0.0cm][0.15cm]{\scriptsize{0 (sym)}}      \\
\hline
\raisebox{-0.05cm}[0.0cm][0.15cm]{\scriptsize{3x3x3}} & \raisebox{-0.05cm}[0.0cm][0.15cm]{\scriptsize{none}}     & \raisebox{-0.05cm}[0.0cm][0.15cm]{\scriptsize{B3LYP/3-21G}} &     \raisebox{-0.05cm}[0.0cm][0.15cm]{\scriptsize{$4.20$}}     &     \raisebox{-0.05cm}[0.0cm][0.15cm]{\scriptsize{0 (sym)}}      \\
\hline
\hline
\raisebox{-0.05cm}[0.0cm][0.15cm]{\scriptsize{3x3 (tilt)}}    & \raisebox{-0.05cm}[0.0cm][0.15cm]{\scriptsize{(1,1,dr)}} & \raisebox{-0.05cm}[0.0cm][0.15cm]{\scriptsize{B3LYP/6-31G}} & \raisebox{-0.05cm}[0.0cm][0.15cm]{\scriptsize{$3.40$}} & \raisebox{-0.05cm}[0.0cm][0.15cm]{\scriptsize{$0.08$}} \\
\hline
\raisebox{-0.05cm}[0.0cm][0.15cm]{\scriptsize{3x3$^{(a)}$}}    & \raisebox{-0.05cm}[0.0cm][0.15cm]{\scriptsize{(1,-1,dl)(1,1,ul)}} & \raisebox{-0.05cm}[0.0cm][0.15cm]{\scriptsize{B3LYP/3-21G}} &\raisebox{-0.05cm}[0.0cm][0.15cm]{\scriptsize{$2.86$}}& \raisebox{-0.05cm}[0.0cm][0.15cm]{\scriptsize{$0.03$}} \\
%         & \scriptsize{(1,1,ul)}  &       &            &               \\
\hline
\raisebox{-0.05cm}[0.0cm][0.15cm]{\scriptsize{3x3$^{(b)}$}}    & \raisebox{-0.05cm}[0.0cm][0.15cm]{\scriptsize{(1,-1,dr)(1,1,ul)}} & \raisebox{-0.05cm}[0.0cm][0.15cm]{\scriptsize{MP2/6-311G}} &\raisebox{-0.05cm}[0.0cm][0.15cm]{\scriptsize{$1.80$}}& \raisebox{-0.05cm}[0.0cm][0.15cm]{\scriptsize{$0.003$}} \\
%         & \scriptsize{(1,1,ul)}  &       &            &               \\
\hline
\raisebox{-0.05cm}[0.0cm][0.15cm]{\scriptsize{3x3x3$^{(c)}$}} & \raisebox{-0.05cm}[0.0cm][0.15cm]{\scriptsize{(0,-1,1,r)(1,-1,0,f)}}     & \raisebox{-0.05cm}[0.0cm][0.15cm]{\scriptsize{HF/6-31+G}} &     \raisebox{-0.05cm}[0.0cm][0.15cm]{\scriptsize{2.15}}     &     \raisebox{-0.05cm}[0.0cm][0.15cm]{\scriptsize{0.04}}      \\
%         & \scriptsize{(1,-1,0,f)}  &       &            &               \\
\hline
\raisebox{-0.05cm}[0.0cm][0.15cm]{\scriptsize{3x3x3 (tilt)}} & \raisebox{-0.05cm}[0.0cm][0.15cm]{\scriptsize{(1,1,dl)}}     & \raisebox{-0.05cm}[0.0cm][0.15cm]{\scriptsize{B3LYP/3-21G}} &     \raisebox{-0.05cm}[0.0cm][0.15cm]{\scriptsize{n.a.}}     &     \raisebox{-0.05cm}[0.0cm][0.15cm]{\scriptsize{0.14}}      \\
\hline
\raisebox{-0.05cm}[0.0cm][0.15cm]{\scriptsize{5x5$^{(d)}$}}& \raisebox{-0.05cm}[0.0cm][0.15cm]{\scriptsize{(0,-2,r)}} & \raisebox{-0.05cm}[0.0cm][0.15cm]{\scriptsize{B3LYP/6-31G}} &  \raisebox{-0.05cm}[0.0cm][0.15cm]{\scriptsize{$4.70$}}   &  \raisebox{-0.05cm}[0.0cm][0.15cm]{\scriptsize{$0.04$}}  \\
\hline
\raisebox{-0.05cm}[0.0cm][0.15cm]{\scriptsize{5x5}} & \raisebox{-0.05cm}[0.0cm][0.15cm]{\scriptsize{(-1,-1,r)}}     & \raisebox{-0.05cm}[0.0cm][0.15cm]{\scriptsize{B3LYP/3-21G}} & \raisebox{-0.05cm}[0.0cm][0.15cm]{\scriptsize{$2.10$}} & \raisebox{-0.05cm}[0.0cm][0.15cm]{\scriptsize{$0.04$}} \\
\hline
\raisebox{-0.05cm}[0.0cm][0.15cm]{\scriptsize{5x5}}  & \raisebox{-0.05cm}[0.0cm][0.15cm]{\scriptsize{(2,0,d)}} & \raisebox{-0.05cm}[0.0cm][0.15cm]{\scriptsize{B3LYP/3-21G}} & \raisebox{-0.05cm}[0.0cm][0.15cm]{\scriptsize{$2.40$}}     &  \raisebox{-0.05cm}[0.0cm][0.15cm]{\scriptsize{$0.11$}}  \\
\hline
\raisebox{-0.05cm}[0.0cm][0.15cm]{\scriptsize{7x7$^{(e)}$}}& \raisebox{-0.05cm}[0.0cm][0.15cm]{\scriptsize{\emph{(see caption)}}} & \raisebox{-0.05cm}[0.0cm][0.15cm]{\scriptsize{HFS/STO-3G}} &  \raisebox{-0.05cm}[0.0cm][0.15cm]{\scriptsize{$10-20$}}   &  \raisebox{-0.05cm}[0.0cm][0.15cm]{\scriptsize{$0.1-0.2$}}  \\
\hline
\end{tabular}
\caption{Some estimations, in eV, for $\gamma_{\rm s}$ and $\gamma_{\rm w}$.
Absolute values are given, since the TLS orientation is arbitrary.
Except for fragments (b-d), all possible TLSs
were calculated, as shown for fragment (a) in table~\ref{tablesmall}, and only
the highest values are displayed here. In the absence of extra impurities,
$\gamma_{\rm w}=0$ for symmetry reasons.
For fragments (b-d), only one $\tau$-TLS and one $S$-TLS were chosen, and
linearity of the energy response was tested as shown in
Fig.~\ref{relaxnewfigure}. For these fragments the values for $\gamma_{\rm w}$
should be considered as lower bounds. tr.ph. stands for "transverse phonon".
Fragment (e) sums up three calculations
with a minimal basis set, with up to four impurities at positions
(0,-2,r)(-2,2,r)(2,0,u),(2,2,d); these results should not be taken on equal
footing with the rest of the table.}
\label{tablegamma}
\end{table}

{\it Results} ---
Fig.~\ref{relaxnewfigure} illustrates some tests of the range of linearity.
One can see that the results are essentially the same for different fragments
and levels of calculations: the first order approximation is very accurate at
least for phonon amplitudes of one or two percent of the interatomic spacing.
On a 3x3 fragment with a central CN$^-$ impurity, at MP2/6-31+G level,
Fig.~\ref{relaxnewfigure} shows the estimation of $\eta=0.6eV$.  The same
conditions yield a comparable $\eta=0.9eV$ for a Cl$^-$ impurity.  In all cases
a noticeable second-order correction of the order of 5eV can be fitted.

The central result of this Letter, reported in table~\ref{tablegamma}, is the
calculation of $\gamma_{\rm s} \simeq 3$eV and $0 \leq \gamma_{\rm w}\leq 0.15$eV.
The finite size of our samples, the quality of our calculation methods, and
differences between planar and cubic samples, all lead to some variance in the parameters.
Yet, the strength of our results lies in the fact that our estimations
are fairly consistent in their order of magnitude for very different lattice fragments and a variety of levels of calculation. This is true for additional calculations, e.g. for a non-central CN$^-$ impurity,
not reported here. The calculations using a
minimal basis set serve to discard a correlation between $\gamma_{\rm w}$ or
$\gamma_{\rm s}$ and the number of impurities.

One should point out that within a given sample and a given level of calculation,
the variance in the values of $\gamma_{\rm s}$ between different orientations is a
result of the small elastic deviations from symmetry, and are therefore a
factor of $g$ smaller than its typical value. With regard to $\gamma_{\rm w}$,
its values are dictated by the aforementioned deviation from local inversion
symmetry. Thus, the distribution of all possible estimations of its values is
peaked at zero, with a variance which equals the typical value. This is displayed
in table~\ref{tablesmall}, for all TLSs in fragment (a). It can also be seen in
the main panel of Fig.~\ref{relaxnewfigure}, where the particular $\tau$-TLS chosen for
fragment (b) experiences a very symmetric environment.  Note that for each
particular combination of fragment and calculation method, the highest
$\gamma_{\rm w}$ values obtained among the four TLS-phonon combinations are
denoted as our estimate values for $\gamma_{\rm w}$ in Table \ref{tablegamma}.
These values are the most relevant for our purposes, as they are expected to be
the best predictors for a real system with many impurities.

\begin{table}[!t]
\begin{tabular}{ccc}
\begin{tabular}{|c|c|}
\hline
&$\gamma_{\rm s}$\\
\hline
\raisebox{-0.1cm}[0.0cm][0.25cm]{$\Delta^{ul}_h$}&\raisebox{-0.1cm}[0.0cm][0.25cm]{ 2.84}\\
\hline
\raisebox{-0.1cm}[0.0cm][0.25cm]{$\Delta^{dr}_h$}&\raisebox{-0.1cm}[0.0cm][0.25cm]{ 2.82}\\
\hline
\raisebox{-0.1cm}[0.0cm][0.25cm]{$\Delta^{ul}_v$}&\raisebox{-0.1cm}[0.0cm][0.25cm]{ {\bf 2.86}}\\
\hline
\raisebox{-0.1cm}[0.0cm][0.25cm]{$\Delta^{dr}_v$}&\raisebox{-0.1cm}[0.0cm][0.25cm]{ 2.83}\\
\hline
\end{tabular}
&
\hspace{1.0cm}
&
\begin{tabular}{|c|c|}
\hline
&$\gamma_{\rm w}$\\
\hline
\raisebox{-0.1cm}[0.0cm][0.25cm]{$\Delta^{ud}_h$}&\raisebox{-0.1cm}[0.0cm][0.25cm]{ 0.0036}\\
\hline
\raisebox{-0.1cm}[0.0cm][0.25cm]{$\Delta^{lr}_h$}&\raisebox{-0.1cm}[0.0cm][0.25cm]{ 0.0070}\\
\hline
\raisebox{-0.1cm}[0.0cm][0.25cm]{$\Delta^{ud}_v$}&\raisebox{-0.1cm}[0.0cm][0.25cm]{ 0.0001}\\
\hline
\raisebox{-0.1cm}[0.0cm][0.25cm]{$\Delta^{lr}_v$}&\raisebox{-0.1cm}[0.0cm][0.25cm]{ {\bf 0.0273}}\\
\hline
\end{tabular}
\end{tabular}
\caption{Values in eV, summary of results for fragment (a) on
Table~\ref{tablegamma}.  Shorthand
$\Delta^{ud}_h=(E^{up}_{horizontal}-E^{up})-(E^{down}_{horizontal}-E^{down})$
has been used for clarity.}
\label{tablesmall}
\end{table}

A noteworthy complication is presented by the in-plane diagonal orientations of
the TLS, because of the small energy difference with the in-space diagonal
states in the real system.  Depending on the fragment, impurities and
calculation method, the relative order of stability changes and the energy of
one or more of the "axial" orientations rises above the most stable "diagonal"
orientation. Some examples of these orientations, denoted as \{dl, ul, dr,
ur\}, are shown in table~\ref{tablegamma}. As illustrated in Figs.~\ref{TLS}
and~\ref{relax}, in the perimeter of the fragment, where the CN$^-$ suffer from
intense border effects, we even find intermediate orientations.  In the cases
where the central TLS is affected by this problem -noted in
Table~\ref{tablegamma} as "tilt"- the extraction of the parameters can be
technically more difficult, but there is no fundamental physical difference
between axial and diagonal orientations as in both cases there is a clear
distinction between S-TLSs and $\tau$-TLSs.\emph{}

{\it Summary} ---
Our numerical calculations confirm qualitatively and quantitatively the results
of Ref.\cite{SS08b}; the existence of weak and strong interacting TLSs in
disordered solids, and the corresponding strength of their interaction
with the phonon field. As TLSs in KBr:CN were experimentally measured
to have a coupling constant of $\gamma \approx 0.12-0.18$eV with the phonon
field, our calculations also verify that it is
indeed the weak interacting $\tau$ TLSs which are the relevant TLSs at low
temperatures, dictating the universal behaviour. Thus, we are able to clearly
identify the relevant TLSs in this particular system.  In Ref.\cite{SS08b} the
plausibility that the same mechanism dictates universality in amorphous solids
was argued for. The verification of this argument requires the identification
of nearly inversion symmetric TLSs in amorphous solids, and the calculation of
their coupling to the phonon field.

We would like to thank Ariel Amir, Lior Kronik, Juan Jos\'e Serrano, Jos\'e
S\'anchez-Mar\'in and Nicolas Suaud for useful discussions.
A.G.A. acknowledges funding by call FP7-PEOPLE-2007-4-1-IOF, project PIOF-GA-2008-219514. 
M.S. acknowledges financial support from the ISF.

\end{document}